\documentclass[preprintnumbers, prd, floatfix, twocolumn, preprintnumbers, letterpaper, superscriptaddress,nofootinbib]{revtex4}
\usepackage{amsfonts}
\usepackage{amsmath}
\usepackage{amssymb,epsf}
\usepackage{latexsym}
\usepackage{graphicx,epsfig}
\usepackage{amssymb}
\usepackage{subfigure}
\usepackage[colorlinks=true,citecolor=blue,linkcolor=blue,urlcolor=black]{hyperref}
\usepackage{epstopdf}

\begin{document}

\title{Comment on "Insight into the Microscopic Structure of an AdS Black Hole from a Thermodynamical Phase Transition"}
\author{M. Kord Zangeneh}
\email{mkzangeneh@shirazu.ac.ir}
\affiliation{Physics Department and Biruni Observatory, Shiraz University, Shiraz 71454,
Iran}
\author{A. Dehyadegari}
\affiliation{Physics Department and Biruni Observatory, Shiraz University, Shiraz 71454,
Iran}
\author{A. Sheykhi}
\email{asheykhi@shirazu.ac.ir}
\affiliation{Physics Department and Biruni Observatory, Shiraz University, Shiraz 71454,
Iran}
\affiliation{Research Institute for Astronomy and Astrophysics of Maragha (RIAAM), P.O.
Box 55134-441, Maragha, Iran}

\begin{abstract}
Thermodynamic geometry analysis of interesting Letter [Phys. Rev. Lett. \textbf{115},
111302 (2015)] for charged AdS black holes, which is based on studying the
Ruppeiner invariant behavior is not correct and the authors made a
mistake in calculating this quantity. In the present Letter, we address the
correct Ruppeiner scalar curvature and reveal the correct possible microscopic
properties of $4$-dimensional charged AdS black holes arise from it. Some of
these properties have not been discussed in pointed out Letter.
\end{abstract}

\maketitle

\textit{Introduction.---}In their interesting Letter \cite{wei}, Shao-Wen
Wei and Yu-Xiao Liu have introduced the number density of the black hole
molecules as a measure for microscopic degrees of freedom of the black hole.
Based on this, they have figured out some microscopic properties of the $4$%
-dimensional charged AdS black hole as an example relying on the
thermodynamic phase transition and thermodynamic geometry, specially the
behavior of the Ricci scalar ($R$) of Ruppeiner geometry \cite{Rup0}. At
first glance, the obtained Ricci scalar seems surprising since shows no
divergency as one usually expects for black holes \cite{Rup1}. This
motivates us to check whether the obtained Ricci scalar is correct. We
observed that Ricci scalar is not correct as we guessed and therefore
discussions and insights about microscopic structure of charged AdS black
holes relying on this should be revised. It is well-known that the main tool
for studying microscopic structure of a black hole is thermodynamic geometry
specially the size and sign of the Ricci scalar of Ruppeiner geometry
(Ruppeiner invariant) \cite{critical,measure,force}.

In this Letter, we address the correct Ruppeiner invariant of the $4$%
-dimensional charged AdS black holes and disclose the correct properties of
the microscopic structure of these black holes. Furthermore, we find that
the positivity of black hole's horizon temperature ($T$) implies an upper
bound on number density of the black hole molecules, the fact that has not
been pointed out in \cite{wei}. Finally, we depict $R-T$ diagram in order to
get more insights about microscopic structure of $4$-dimensional charged AdS
black holes.

\textit{Thermodynamic geometry of AdS black holes.---}Since Ricci scalar is
a thermodynamic invariant, the Ruppeiner geometry defined in ($M$, $P$)
space by taking entropy $S$ as thermodynamic potential, can be rewritten in
the Weinhold energy form \cite{Rup1}%
\begin{equation}
g_{\alpha \beta }=\frac{1}{T}\frac{\partial ^{2}M}{\partial X^{\alpha
}\partial X^{\beta }},
\end{equation}%
in ($S$, $P$) space in order to calculate it. Using the mass, $%
M=Q^{2}/2r_{h}+r_{h}/2+4\pi r_{h}^{3}P/3$, and temperature $T=1/4\pi
r_{h}+2r_{h}P-Q^{2}/4\pi r_{h}^{3}$ of $4$-dimensional charged AdS black
hole, where $r_{h}=\sqrt{S/\pi }$, one can obtain the Ruppeiner scalar
curvature (Ricci scalar) as%
\begin{equation}
R=\frac{2\pi Q^{2}-S}{8PS^{3}+S^{2}-\pi SQ^{2}}.
\end{equation}%
Furthermore, $R$ can be rewritten in terms of number density $n=1/2r_{h}$ as%
\begin{equation}
R=\frac{1}{3\pi }\,{\frac{\left( n/n_{c}\right) ^{6}-3\left( n/n_{c}\right)
^{4}}{\,-\left( n/n_{c}\right) ^{4}+6\left( n/n_{c}\right) ^{2}+3\left(
P/P_{c}\right) },}  \label{Ric}
\end{equation}%
where we have set $Q=1$, and ($n_{c}$, $P_{c}$)$=$($1/2\sqrt{6}$, $1/96\pi $%
).

Some interesting results can be concluded from Eq. (\ref{Ric}). We know that
the sign of $R$ determines the kind of intermolecular interaction for the
thermodynamic system \cite{Osh}. Positivity (negativity) of $R$ shows the
dominance of repulsive (attractive) intermolecular interaction while $R=0$
indicates there is no interaction as in the case of classical ideal gas.
Moreover, the diverging behavior of $R$ may be occured either at absolute
zero temperature\textbf{\ }or at critical points \cite{Rup1}.

Let us first examine the cases where black hole micromolecules behave as
classical ideal gas ones. This occurs in two cases where $R=0$. The first
case is when $n=\sqrt{3}n_{c}$ provided the denominator of (\ref{Ric}) is
not zero. The second case corresponds to very large black holes in a fixed
AdS space when $n/n_{c}\sim 0$. It is notable to mention that in the latter
case, $R$ approaches to zero while in former it is exactly zero. Now, let us
turn to investigate the diverging behavior of Ruppeiner scalar curvature.
The divergency of $R$ took place at absolute zero temperature $T=0$ where
the black hole is extremal. We can also gain insight about microscopic
properties of charged AdS black hole in different cases by using Eq. (\ref%
{Ric}). For $n/n_{c}<1$, we encounter a negative $R$. This means that for
large black holes, the possible molecules attract each other. In the case of 
$n/n_{c}>1$, the sign of Ricci scalar (\ref{Ric}) depends on the value of $%
P/P_{c}$ and can be positive, negative, zero or infinity (two latter cases
have been investigated above). In this case, table \ref{tab} shows that we
have negative scalar curvature (attractive intermolecular interaction) for $%
1<n/n_{c}<\sqrt{3}$ while $R>0$ (repulsive intermolecular interaction) for $%
\sqrt{3}<n/n_{c}<\sqrt{n_{0}}$ where $n_{0}=3+\sqrt{9+3P/P_{c}}$. Table \ref%
{tab} also shows that $n/n_{c}>\sqrt{n_{0}}$ is excluded because temperature
is negative in this case. 
\begin{table}[h]
\caption{The allowed ranges of $(n/n_{c})>1$.}
\label{tab}%
\begin{tabular}{c|c|c|c|}
\cline{2-4}
& $n/n_{c}<\sqrt{3}$ & $\sqrt{3}<n/n_{c}<\sqrt{n_{0}}$ & $n/n_{c}>\sqrt{n_{0}%
}$ \\ \hline
\multicolumn{1}{|c|}{$R$} & $-$ & $+$ & $-$ \\ \hline
\multicolumn{1}{|c|}{$T$} & $+$ & $+$ & $-$ \\ \hline
\multicolumn{1}{|c|}{validity} & allowed & allowed & not allowed \\ \hline
\end{tabular}%
\end{table}

\begin{figure}[t]
\includegraphics[width=.46\textwidth]{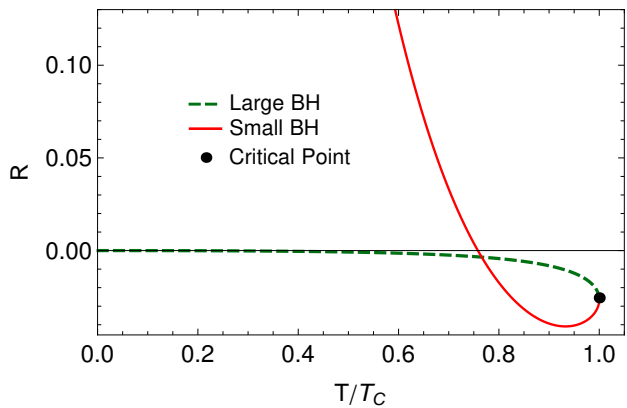}\qquad
\caption{The behavior of Ricci scalar $R$ along the coexistence curve for
small and large BHs.}
\label{fig1}
\end{figure}

The standard diagram which discloses the microscopic properties along
coexistence curve and presents more insights into the microscopic structure
of black holes is $R-T$ diagram \cite{critical,Rup1,Rup2}. We depict this
diagram in Fig. \ref{fig1}. This figure shows that there exist a point under
critical point where there is no gap in $R$ and thus small and large black
holes have similar effective attractive interactions in this point exactly
like critical point. Furthermore, Fig. \ref{fig1} reveals that extremal
small black holes behave like fermion gas near zero temperature \cite%
{measure} while extremal large black holes resemble ideal gas \cite{Rup3}.
Moreover, there are attractive interactions between large black holes
molecules while the type of molecular interactions change at $T=0.76T_{c}$
for small black holes. 

\textit{Summary.---}In last part of interesting Letter \cite{wei}, the Ricci
scalar of Ruppeiner geometry has been calculated in order to study some
thermodynamical and microscopical properties of $4$-dimensional charged AdS
black holes. The Ricci scalar obtained in \cite{wei} is not correct. In this
Letter, we calculated the correct Ruppeiner scalar curvature ($R$) of $4$%
-dimensional charged AdS black holes and revealed the correct
thermodynamical and microscopical properties of these solutions. We also
showed that the positivity of black hole temperature impose an upper bound
on number density ($n$) of the black hole molecules which has been
introduced as a measure for microscopic degrees of freedom of the black
holes in \cite{wei}. Finally, we depicted $R-T$ diagram in order to get more
insights about microscopic structures of $4$-dimensional charged AdS black
holes. This diagram shows that in addition to critical point there is
another point at which the effective attractive interactions for small and
large black holes are the same. Moreover, we concluded from this diagram
that extremal small and large black holes behave like fermion gas near zero
temperature and ideal gas respectively. We also found that the type of
molecular interactions for small black holes change at $T=0.76T_{c}$.

\end{document}